\documentclass[nato,noid,numreferences]{crckbked}
\usepackage{bbm}

\usepackage{graphicx}
\usepackage{epsf}
\newcommand{\cA}{{\cal A}}

\newcommand{\be}{\begin{eqnarray}}
\newcommand{\ee}{\end{eqnarray}}
\newcommand{\hk}{\hspace{0.1cm}}

\newcommand{\rk}{\right)}
\newcommand{\lk}{\left(}
\newcommand{\rke}{\right]}
\newcommand{\lke}{\left[}

\begin{document}
% The \begin{document} command comes before the \begin{opening}
% command.

\begin{opening}
\title{Topology of Center Vortices\thanks{Talk given at the NATO workshop on ``Confinement,
Topology, and other Non-Perturbative Aspects of QCD'', Stara Lesna, Slovakia,
January 21-27, 2000.}}

%\subtitle{Basic Instructions}
% Uncomment if you want to give a subtitle.

% You can split the title and subtitle by putting 
% two backslashes at the appropriate place. 

\author{H. REINHARDT\thanks{Supported by DFG
under grant number DFG-Re856/4-1.}}
\institute{Institut f\"ur Theoretische Physik\\
           Universit\"at T\"ubingen\\
	   Auf der Morgenstelle 14\\
	   D-72076 T\"ubingen}
% If there are more authors at one institute, you should first
% use \author{...} for each author followed by \institute{...}.
% Put 
% \author{} 
 
% before the last author.

%\author{and}
%\institute{}

%\author{Second AUTHOR}
%\institute{Affiliation\\
           %Institute address}

\begin{abstract}
In this talk I study the topology of mathematically idealised center vortices, defined in a gauge invariant way
as closed (infinitely thin) flux surfaces (in D=4 dimensions) which contribute
the $n^{th}$ power of a
non-trivial center element to Wilson loops when they are n-foldly linked to
the latter. In ordinary 3-space generic center vortices represent closed
magnetic flux loops which evolve in time. I show that
the topological charge of such a time-dependent vortex loop can be entirely
expressed by the temporal changes of its writhing number.
\end{abstract}

\end{opening}
\section{Introduction}
\label{dummy}

The vortex picture of the Yang-Mills vacuum gives an appealing explanation of
confinement. This picture introduced already in the late 70s
\cite{[Hooft]} has only
recently received strong support from lattice calculations performed in the
so-called maximum center gauge \cite{[Del2]} where one fixes only the coset $G/Z$,
but leaves the center $Z$ of the gauge group $G$ unfixed. In this gauge the
identification of center vortices can be easily accomplished by means of the 
so-called center projection, which consists of replacing each link by its closest
center element. The vortex content obtained in this manner is a physical property of the
gauge ensemble \cite{[Lang]} 
and produces virtually the full string tension
\cite{[Ldel]}. Furthermore, the string tension disappears when the center
vortices are removed from the
Yang-Mills ensemble \cite{[Del2]}. This property of center dominance of the string tension
survives at finite temperature and the deconfinement phase transition can be
understood in a 3-dimensional slice at a fixed spatial coordinate as a
transition from a percolated vortex phase to a phase in which vortices cease to
percolate \cite{[Klang]}. Furthermore, by calculating the free energy of center vortices it has
been shown that the center vortices condense in the confinement
phase \cite{[Kova]}. It has also
been found on the lattice that if the center vortices are removed from the
gauge ensemble, chiral symmetry breaking disappears and all field configurations
belong to the topologically trivial sector \cite{[Forc]}. 
Thus center vortices might simultaneously provide a description of confinement
and spontaneous breaking of chiral symmetry, in accord with the lattice observation that the
deconfinement phase transition and the restoration of chiral symmetry occur at the
same temperature. Usually, spontaneous breaking of chiral symmetry is attributed to
instantons \cite{[YY]}, which, however, do not explain confinement. These
topologically non-trivial configurations give rise to quark zero modes localized
at the center of the instantons. In an ensemble of (anti-) instantons these zero
modes start overlapping and form a quasi-continuous band of states near zero
virtually, which by the Banks-Casher relation gives rise to a quark consdensate,
the order parameter of spontaneous breaking of chiral symmetry. This phenomenon
is obviously related to the topological properties of gauge fields. Furthermore,
center vortices seem to acount also for the topological susceptibility
\cite{[Ber2]}

\medskip

In the present paper I
study the topology of generic center vortices, which represent (in general time-dependent) closed magnetic flux loops, and express their topological charge in
terms of the topological properties of these loops. I will show that
the topological charge of generic center vortices is given by the temporal change of
the writhing number of the magnetic flux loops. My talk is mainly based on
ref. \cite{[Reinhx]}.
\medskip

\section{Center vortices in continuum Yang-Mills theory}

In D-dimensional continuum Yang-Mills theory center vortices are localised gauge 
field configurations $A_\mu (x)$ whose flux is concentrated on  $D=2$ dimensional closed hypersurfaces 
$\partial \Sigma$, and which produce a Wilson loop

\be
\label{G1}P e^{- \oint\limits_C dx_\mu A_\mu (\partial \Sigma)} = 
Z^{L (C, \partial \Sigma)} \hk ,
\ee

where $Z$ denotes a non-trivial center element of the gauge group and 
$ L (C, \partial \Sigma)$ is the linking number between the (large) Wilson loop C and the closed vortex
hypersurface $\partial \Sigma$. 
For the present considerations, where I
concentrate on the topological properties of center vortices, it is sufficient to
consider mathematically idealised center vortices whose flux 
${\cal F}_{\mu \nu}
\lk \partial \Sigma \rk $ lives entirely on the closed hypersurface $\partial
\Sigma $ 

\be
\label{F2}{\cal F} _{\mu \nu} \lk \partial \Sigma ,x \rk = E \int\limits_{\partial \Sigma} d^{D-2}
\tilde{\sigma}_{\mu \nu} \delta^{(D)} \lk x - \bar{x} (\sigma) \rk,
\ee

where $\bar x _\mu \lk \sigma \rk $ is a parametrization of the vortex surface
$\partial \Sigma $ and $E$ denotes a co-weight of the gauge group satisfying
$exp \lk -E \rk = Z $.

\medskip

Whether the flux of a center vortex (\ref{F2}) is electric or magnetic, or both depends on
the position of the ($D - 2$)-dimensional vortex surface $\partial \Sigma$ in 
$D$-dimensional space.

\section{The topological charge of center vortices in terms of intersection points}

The topology of gauge fields is characterised by the topological charge
(Pontryagin index)

\be
\label{G25}\nu [A] = - \frac{1}{16 \pi^2}  \int d^4 x t r F_{\mu \nu} \tilde{F}_{\mu \nu}
\hk .
\ee

For center vortices with field strength (\ref{F2}) one finds \cite{[Eng]} (see
also refs. \cite{[Reinh2]}, \cite{[Corn]})

\be
\label{G27}\nu [ \cA (\Sigma) ] = \frac{1}{4} I \lk \partial \Sigma , \partial
\Sigma \rk \hk ,
\ee

where $ I \lk S_1, S_2 \rk = \frac{1}{2} \int\limits_{S_1} d \sigma_{\mu \nu}
\int\limits_{S_2} d
\tilde{\sigma}'_{\mu \nu} \delta^{(4)}\lk \bar{x} (\sigma) - \bar{x} (\sigma')
\rk $ is the oriented intersection number of two 2-dimensional (in general open)
surfaces $S_1, S_2$ in $R^4$. Generically, two 2-dimensional surfaces intersect
in $R^4$ at isolated points. The self-intersection number $I \lk \partial \Sigma
, \partial \Sigma \rk $ recieves contributions from two types of singular
points: 

\begin{figure}[ht]
\centerline{
\epsfysize=4.0cm
\epsffile{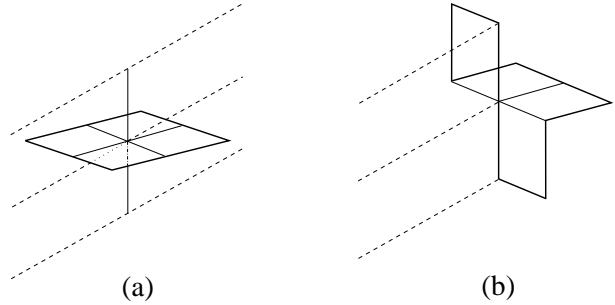}
}
\caption{Illustration of (a) a transversal intersection point and (b) a
  twisting point. The dashed lines indicate the fourth dimension (time
  direction).}
    \label{fig1}
\end{figure}

(i) Transversal intersection points, arising from the intersection of two different
surface patches see fig. \ref{fig1} (a), and (ii) twisting points occuring on a single surface patch
twisting around a point in such a way to produce four linearly independent
tangent vectors, see fig. \ref{fig1} (b).
Transversal intersection points yield a contribution $\pm$ 2 to the oriented
intersection number $ I \lk \partial \Sigma , \partial \Sigma \rk $, 
where the sign depends on the relative orientation of the two
intersecting surface pieces. 
Twisting points yield always contributions of module smaller than 2. For closed oriented surfaces the oriented
self-intersection number $I \lk \partial \Sigma , \partial \Sigma \rk $
vanishes. Center vortices
with non-zero topological charge consist of open differently oriented 
surface patches joined by magnetic monopole loops.
In fact, the topological charge of center vortices (\ref{G27}) can be expressed
as \cite{[Eng]} $ \nu = \frac{1}{4} L \lk C , \partial \Sigma \rk $
where $L \lk C , \partial \Sigma \rk $ is the linking number between the center
vortex surface $\partial \Sigma $ and the magnetic monopole loops $C$ on it. 

\begin{figure}[ht]
\centerline{
\epsfysize=6.0cm
\epsffile{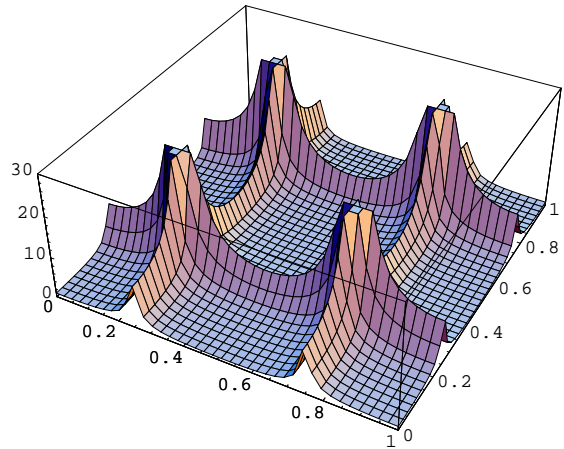}
}
  \caption{ The probability density of the zero mode of quarks in the background
  of four intersecting planar center vortices \cite{[zz]}. In the two-dimensional cut of the
  4-dimensional universe shown in the figure the vortices appear as intersecting
  lines. }
    \label{fig2}
\end{figure}

By the Athiya-Singer-Index theorem $ \nu = N_L - N_R $,
a non-zero topological charge $\nu$ is connected to the difference between the
numbers $N_{L/R}$ of left and right handed
quark-zero modes. Figure \ref{fig2} shows the probabiltiy density of the zero modes of
the quarks moving in the backround of two pairs of intersecting center vortices
on the 4-dimensional torous \cite{[zz]}. As one observes, the quark-zero modes are
concentrated on the center vortex sheets and are in particular localized at the
intersection points, the spots of topological charge $\nu = \frac{1}{2}$. If the
quark-zero modes dominate the quark propagator the quarks will travel along the
center vortex sheets and can move from one vortex to an other through the
intersection points. Since the center vortices percolate in the QCD vacuum we
expect also the percolation of the quark trajectories, which will eventually
result in a condensation of the quarks. 

\section{Topology of generic center vortices}

Generically at a fixed time $t$ a center vortex $\partial \Sigma$ represents a
closed magnetic flux loop $C(t)$. For such magnetic flux loops  topological
charge can be expressed as \cite{[Reinhx]} 

\be
\label{G83}\nu = \frac {1}{4} \int dt \partial_t W \lk C (t) \rk ,
\ee 

where $W (C)$ denotes the writhing number of $C$, which is defined as the 
coincidence limit $W (C) = L (C,C)$ of the Gaussian linking number

\be
\label{G82}
L \lk C_1 ,C_2 \rk & = & \frac{1}{4\pi} \oint \limits_{C_1} dx_i \oint\limits_{C_2} d x'_j
\varepsilon_{i j k} \frac {x_{k} - x'_{k}}{\left | \vec{x} -\vec{x'}\right |^3}.
\ee

If the writhing number changes continuously during the whole time evolution say
from an initial time $t_i$ to a final time $t_f$ (i.e. $W \lk C(t)\rk$ is
a differentiable function of time) the topological charge is given by 
$ \nu = \frac{1}{4} \lk W (t_f ) - W (t_i )\rk $.
However, the writhing number $W (t)$ may change in a discontinuous way, e.g. when two line
segments of the vortex loop intersect (see below).
If we denote by $\bar{t}_k , k= 1,2,..; t_i < \bar{t}_k < t_f$ the
intermediate time instants where $W(t)$ jumps by a finite amount 
$ \Delta W \lk \bar{t}_k \rk = \lim \limits_{\varepsilon \rightarrow 0} \lke W \lk
\bar{t}_k + \varepsilon \rk - W \lk t_k - \varepsilon \rk \rke $
the complete expression for the topological charge for a generic center
vortex is given by \cite{[Reinhx]}

\be
\label{G86}\nu = \frac{1}{4} \lke W \lk t_f \rk + \sum \limits_k \Delta W \lk \bar{t}_k \rk
- W \lk t_i \rk \rke .
\ee

This relation will be illustrated below by means of an example.

\begin{figure}[ht]
\centerline{
\epsfysize=8.5cm
\epsffile{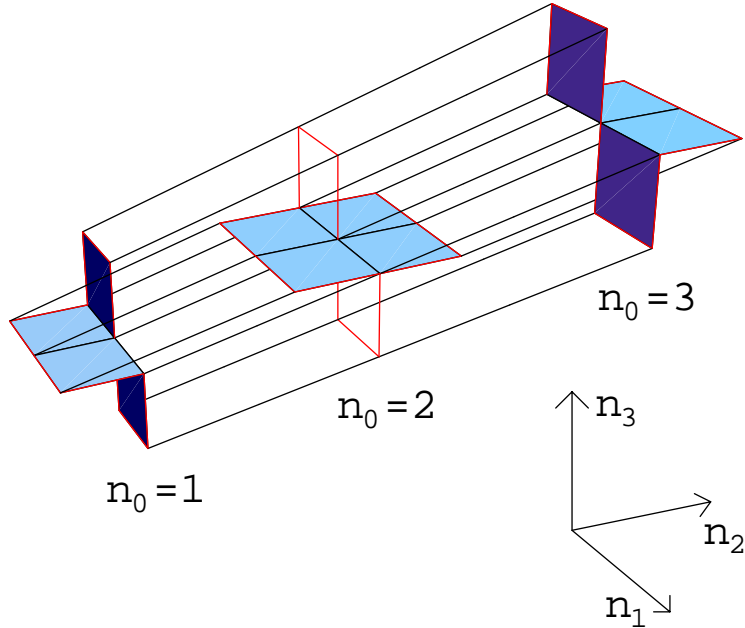}
}
\caption{Sample vortex surface configuration taken from \cite{[Eng2]}. At each lattice time
$t=n_0 a$ (a is the lattice spacing), shaded plaquettes are part of the vortex surface. These
plaquettes are furthermore connected to plaquettes running in time
direction; their location can be inferred most easily by keeping in
mind that each link of the configuration is connected to exactly
two plaquettes (i.e. the surface is closed and contains no intersection
lines). Note that the two non-shaded plaquettes at $n_0 =2$ are {\em not}
part of the vortex; only the two sets of three links bounding them are.
These are slices at $n_0 =2$ of surface segments running in time
direction from $n_0 =1$ through to $n_0 =3$. Sliced at $n_0 =2$, these
surface segments show up as lines. Furthermore, by successively assigning
orientations to all plaquettes, one can convince oneself that the
configuration is orientable. The vortex image was generated by means of a
MATHEMATICA routine provided by R.~Bertle
and M.~Faber.}
\label{fig3}
\end{figure}

\section{The writhing number of center vortex loops}

To illustrate the various singular vortex points, let us consider as an example
the
center vortex configuration shown in fig. \ref{fig3}, \cite{[Eng2]}, which could
arise in a lattice
simulation after center projection \cite{[Ber]}, or in a random vortex
model \cite{[Eng3]}. This vortex surface is orientable
and has various spots of non-zero topological charge: 
There is a transversal intersection point at the intermediate time\footnote{Here the time $t = n_0 a$ 
is quoted in (integer) units $n_0$ of lattice spacing
$a$.} $n_0 =2$
contributing $\frac{1}{2}$ to the topological charge $\nu$. At this time there
are also two twisting points at the front and back edges of the configuration,
each contributing $- \frac{1}{8}$ to $\nu$. Further twisting points occur at the initial $(n_0 =1)$ and
final $(n_0 =3)$ times, each contributing $-\frac{1}{8}$ to $\nu$, so that the
total topological charge vanishes  for this vortex configuration $(\nu = 0)$.

\begin{figure}
\centerline{
\epsfysize=4.0cm
\epsffile{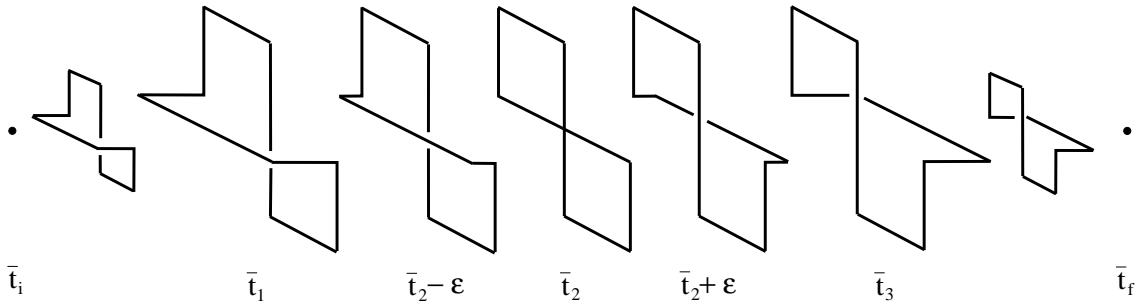}
}
  \caption{Snap shots at characteristic time instants of the continuum center 
    vortex loop whose lattice realization is shown in fig.
     \ref{fig3}.}
\label{fig4}
\end{figure}

Let us now interpret the same configuration as a time dependent vortex loop
in ordinary 3-dimensional space (as in a movie-show) eliminating lattice
artifacts due to the use of a discretised time \cite{[Reinhx]}. Purely spatial vortex patches can be considered as lattice artifacts.
They represent the discrete time step approximation to continuously evolving (in
time) vortex loops. Fig. \ref{fig4} shows the time-evolution of a closed magnetic vortex
loop in ordinary 3-dimensional space which on the 4-dimensional lattice gives rise to the
configuration shown in fig. \ref{fig3}. For simplicity I have kept the cubistic
representation in $D = 3$ space, so that the loops consist of straight
line segments. At an initial time $t = \bar{t}_i$ an infinitesimal
closed vortex loop is generated which then growths up to a time
$t =\bar{t}_1$. Then the long horizontal loop segment moves towards, and at time $t =
\bar{t}_2$ crosses the long vertical loop segment, and continues to move up to a time $t =
\bar{t}_3$. After this time the loop decreases continuously and at the fixed time
$t = \bar{t}_f$ shrinks to a point.

\medskip

For simplicity let us choose $t_i < \bar{t}_i$ (the birth of
the vortex loop) and $t_f > \bar{t}_f$ (the death of the vortex loop). Then 
$W (t_i) = W (t_f) = 0$ since there are no vortex loops at the initial and final
times. In the field 
configuration shown in fig. \ref{fig4} there are discontinuous changes of the vortex
loop, and accordingly of the writhing
number, at the creation (birth) of the vortex loop at $t = \bar{t}_i$, at the
intermediate time $\bar{t}_2$,  where two line segments cross and two lines turn by 180
degrees, and at the annihilation (death) of the vortex loop at $t = \bar{t}_f$. Hence the
topological charge of this configuration is given by (assuming $t_i < \bar{t}_f$
and $t_f > \bar{t}_f$) 

\be
\label{G92}\nu = \frac{1}{4} \lke \Delta W \lk \bar{t}_f \rk + \Delta W \lk \bar{t}_2 \rk +
\Delta W \lk \bar{t}_i\rk \rke
\ee 

For simplicity, let us assume that from its creation at
$\bar{t}_i$ until the time $\bar{t}_1$ the vortex loop does not change its
shape, but merely scales in size. The same will be assumed for the vortex evolution
from $\bar{t}_3$ until its annihilation at $\bar{t}_f$. Then the change of the
writhing number at vortex creation and at annihilation, respectively, is given by
$\Delta W (\bar{t}_i) = W (\bar{t}_1)$ and $\Delta W (\bar{t}_f) = - W
(\bar{t}_3)$, so that we obtain for the topological charge (\ref{G92})

\be
\label{nu}\nu = \frac{1}{4} \lke W (\bar{t}_1) + \Delta W (\bar{t}_2) - W
(\bar{t}_3) \rke .
\ee

The writhing numbers $W (\bar{t}_1) ,  W (\bar{t}_3)$ are explicitly evaluated
in ref. \cite{[Reinhx]} $W (\bar{t}_1) = \frac{1}{2} ,W (\bar{t}_3) = -
\frac{1}{2}$.
\medskip

A further singular change of the vortex loop shown in
fig. \ref{fig4} occurs at the intermediate time $t=\bar{t}_2$. 
At this time the two long line segments 
intersect. The crossing of these two line segments at $t = \bar{t}_2 , \lk
x,y,z \rk = \lk 0,0,0 \rk $ corresponds in D=4 to the transversal intersection point
shown in fig. \ref{fig3} at $n_0 =2$. In fact, in ref. \cite{[Reinhx]} it is shown that the
crossing of these two line segments
gives rise to a change in the writhing number of 
$ \Delta W \lk \bar{t}_2 \rk ^{(i)} = -2 $, which in view of eq.~(\ref{G86}) is in accord with the finding \cite{[Eng]} that a
transversal intersection point contributes $\Delta \nu = \pm \frac{1}{2}$ to the
topological charge.
Furthermore, when the two long loop segments cross the two short horizontal loop segments
at the front and back edges reverse
their directions,  which can be interpreted as twisting these loop segments by an
angle $\pi$ around the $n_1$-axis. In the $D=4$ dimensional lattice realization of the present center vortex shown in
fig. \ref{fig3} these twistings of the vortex loop segments (in $D=3$) by angle $\pi$
correspond to the two twisting points at $n_0 =2$ at the front and back edges of
the configuration. As shown in ref. \cite{[Reinhx]} these two twisting points 
both change the writhing number
by $ \Delta W \lk \bar{t}_2 \rk ^t = \frac{1}{2}$
and hence contribute $\Delta \nu = \frac{1}{8} $ to the topological charge,
again in agreement with the analysis of $\nu$ in $D=4$. As a result we find for
the total change in the writhing number at $t = \bar t _z, \Delta W \lk \bar t
_2 \rk = -2 + \frac{1}{2} + \frac{1}{2} = -1$.
\medskip

In ref. \cite{[Reinhx]} also the twist of the vortex loops was studied. It was
found that transversal intersection points, corresponding to the crossing of 2
full line segments, never change the twist, while twisting points, depending on
the choosen framing, are usually also connected to changes of the twist, which
justifies their name. Finally let us also mention that the description of the
topological charge of center vortices in terms of the temporal changes of the
writhing number of the time-dependent vortex loops ramains also valid for
non-oriented center vortices, i.e. in the presence of magnetic monopole loops. 
\bigskip

{\bf Acknowledgements:}

I thank the organizers, J. Greensite and $\check{S.}$ Olejnik, for bringing us
together to this interesting workshop. Discussions with M. Engelhardt, T. Tok and I. Zahed  are
gratefully acknowledged.


\begin{thebibliography}{99}

\bibitem{[Hooft]}G.`t Hooft, Nucl. Phys. {\bf{B138}} (1978) 1;\\
Y. Aharonov, A. Casher and S. Yankielowicz, Nucl. Phys. {\bf{B146}}
(1978) 256;\\ J. M. Cornwall, Nucl. Phys. {\bf{B157}} (1979) 392\\
G. Mack and V. B. Petkova, Ann. Phys. (NY) {\bf{123}} (1979) 442;\\
G. Mack, Phys. Rev. Lett. {\bf{45}} (1980) 1378;\\
G. Mack and V. B. Petkova, Ann. Phys. (NY) {\bf{125}} (1989) 117;\\
G. Mack, in: {\it{Recent Developments in Gauge Theories}} , eds. G. 't Hooft et
al. (Plenum, New York, 1980); \\ G. Mack and E. Pietarinen, Nucl. Phys.
{\bf{B205}} [FS5] (1982) 141\\
H. B. Nielsen and P. Olesen, Nucl. Phys. {\bf{B160}} (1979) 380;\\ H.
Ambj\o rn and P. Olesen, Nucl. Phys. {\bf{B170}} [FS1] (1980) 60;\\ J.
J. Ambj\o rn and P. Olesen, Nucl. Phys. {\bf{B170}} [FS1] (1980) 265;\\
E. T. Tomboulis, Phys. Rev. {\bf{D 23}} (1981) 2371

\bibitem{[Del2]}Del Debbio, M. Faber, J. Greensite, $\breve S$. Olejnik, Phys.
Rev. {\bf{D55}} (1997) 2298

\bibitem{[Lang]}K. Langfeld, H. Reinhardt, O. Tennert, Phys. Lett.
{\bf{B419}} (1998) 317

\bibitem{[Ldel]}L. Del Debbio. M. Faber, J. Giedt, J. Greensite and $\breve S$ .
Olejnik, Phys. Rev. {\bf{D 58}} (1998) 094501

\bibitem{[Klang]}K. Langfeld, O. Tennert, M. Engelhardt and H. Reinhardt, Phys.
Lett. {\bf{B542}} (1999) 301, M. Engelhardt, K. Langfeld, H. Reinhardt
and O. Tennert, Phys. Rev. {\bf{D61}} (2000) 054504 

\bibitem{[Kova]}T. G. Kovacs, E. T. Tomboulis, Phys. Rev. Lett.
{\bf{85}} (2000) 704

\bibitem{[Forc]}P. de Forcrand and M. D`Elia, Phys. Rev. Lett. {\bf{82}}
(1999) 4582.


\bibitem{[YY]}M. Nowak, M. Rho, I. Zahed, Chiral Nuclear Dynamics, World
Scientific, Singapore, 1996 and references therein

\bibitem{[Ber2]}R. Bertle, M. Engelhardt, M. Faber, Phys. Rev. {\bf{D64}} (2001)
504

\bibitem{[Reinhx]}H. Reinhardt, hep-th/0112215, Nucl. Phys. B, in press.

\bibitem{[Eng]}M. Engelhardt, H. Reinhardt, Nucl. Phys. {\bf{B567}}
(2000) 249

\bibitem{[Reinh2]} H. Reinhardt, M. Engelhardt, Proceedings of the
XVIII Lisbon Autumn School, ``Topology of Strongly Correlated Systems'', Lisbon,
8-13 October, 2000, hep-th/0010031 

\bibitem{[Corn]}J. M. Cornwall, Phys. Rev. {\bf{D61}} (2000) 085012

\bibitem{[Eng3]} M. Engelhardt and H. Reinhardt, Nucl. Phys. {\bf{B585}}
(2000) 591

\bibitem{[Eng2]}M. Engelhardt, Nucl. Phys. {\bf{B585}} (2000) 614

\bibitem{[Reinh]}H. Reinhardt, Nucl. Phys. {\bf{B503}} (1997) 505

\bibitem{[Ber]}R. Bertle, M. Faber, J. Greensite. $\breve S$. Olejnik, JHEP
{\bf{9903}} (1999) 019 

\bibitem{[zz]} H. Reinhardt, O. Schr\"oder, T. Tok, V. Ch. Zhukovsky,
hep-th/0203012

\end{thebibliography}
\end{document}